\date{}
\documentclass[showpacs,superscriptaddress]{revtex4}
\usepackage{graphicx,psfrag,amsmath,amssymb,amsfonts,bbm,latexsym,color,dcolumn,epsf}

\begin{document}

\title{Effective action for $QED_3$ in a region with borders}
\author{C.~D.~Fosco}
\affiliation{Centro At\'omico Bariloche and Instituto Balseiro,
Comisi\'on Nacional de Energ\'\i a At\'omica - R8402AGP Bariloche,
Argentina}
\author{F.~D.~Mazzitelli}
\affiliation{Departamento de F\'\i sica, Facultad de Ciencias
Exactas y Naturales, Universidad de Buenos Aires, Ciudad
Universitaria, Pabell\' on I - 1428 Buenos Aires, Argentina}

\begin{abstract}
\noindent
We study quantum effects due to a Dirac field in $2+1$ dimensions, confined to
a spatial region with a non-trivial boundary, and minimally coupled to an
Abelian gauge field. To that end, we apply a path-integral representation,
which is applied to the evaluation of the Casimir energy and to the study of
the contribution of the boundary modes to the effective action when an
external gauge field is present.  We also implement a large-mass expansion,
deriving results which are, in principle, valid for any geometry. We compare
them with their counterparts obtained from the large-mass `bosonized'
effective theory.
\end{abstract}
\maketitle
\section{Introduction}\label{sec:intro}
The presence of borders drastically modifies the energy spectrum of a quantum
field, by producing a vacuum energy with a non trivial dependence on the
geometry of the borders and the detailed form of the boundary conditions.  The
resulting `Casimir energy' has many interesting physical consequences, ranging
from the existence of forces between uncharged metallic surfaces to
potentially relevant effects in some cosmological scenarios~\cite{review}.

On the other hand, quantum theories in the presence of background fields
naturally arise in many different physical situations, like when considering
the effects of classical gravitational or electromagnetic background fields on
the vacuum persistence amplitude.
Besides, the consideration of `classical' backgrounds is sometimes an
important intermediate step in the context of the functional quantization
approach, whereby one considers (trivial or non trivial) classical
backgrounds.
Those configurations may afterwards be allowed to fluctuate; usually this is
done without modifying either the topology or the boundary conditions of the
classical background.

Quantum fields coupled to background fields and models defined on spaces with
non trivial borders do share some important properties. Indeed, the latter can
sometimes be regarded as a special limit of the former.  Background fields do
of course also modify the energy spectrum in a non trivial way. As a result of
this, the vacuum persistence amplitude, obtained by integrating out the
quantum fields becomes a (usually) complicated functional of the background
field.

$QED$ in $2+1$ dimensions is an interesting arena for the analysis of the
combined effect of boundary conditions and background fields on the quantum
vacuum.
The Casimir energy for massless and massive spinor fields in 2+1
dimensions has been discussed at length, using the zeta function approach
\cite{zeta}. The effect of boundary conditions in the presence of external
fields have also received some attention, in particular in the case of
fermions satisfying MIT boundary conditions on a circle in the presence of a
magnetic flux string~\cite{beneventano}.

In this paper, we shall consider a path integral approach to the computation
of the effective action in the presence of non trivial boundaries and external
fields. This approach, introduced in~\cite{kardar}, has been adapted to the case
of the
electromagnetic field satisfying perfect conductor boundary conditions on the
borders \cite{kardar2}, and successfully applied to the calculation of Casimir
forces in different geometries \cite{recent}. The main idea is to implement
the boundary conditions as delta functions in the functional integral, and to
write them in terms of auxiliary fields living on the boundaries.  Here we
will apply a similar idea to the case of a Dirac field in $2+1$ dimensions. We
will assume that the field is confined into a static spacetime region, and that
is minimally coupled to an Abelian gauge field. We shall obtain a general
formula for the effective action in terms of a non local kernel evaluated on
the boundary. We will then analyze some of its formal properties, applying it
next to the calculation of the Casimir energy and of the contribution of the
borders to the effective action for the gauge field.

The paper is organized as follows. In section~\ref{sec:effect}, we adapt the
method of~\cite{kardar} to the present case. That approach is also used to
understand the issue of gauge invariance, and to calculate the fermion
propagator in the same system.

After studying some general properties of those functional representations, we
apply them, in the following sections, to calculate the effective action under
different approximation schemes and simplifying assumptions. In
section~\ref{sec:casimir}, we consider the Casimir energy for massless Dirac
fermions, which is derived from the effective action with a vanishing gauge
field, for the special geometry of two parallel plates.

In section~\ref{sec:largem}, we evaluate the effective action in a large-$m$
approximation, for the case on an arbitrary external gauge field.  This yields
a contribution coming from the boundary modes, which is local when the mass
tends to infinity.  In this section, we also discuss the same system from a
different point of view: we start from the `dual' or bosonized version of the
Dirac field in the large-mass limit, which is a Chern-Simons action. This
action is then constrained to satisfy the corresponding boundary condition,
which now is a kind of `perfect conductor' boundary condition for the
Chern-Simons gauge field.  We obtain the resulting functional integral for the
boundary modes, and compare with the previous result.

In section~\ref{sec:linear}, we consider the dependence of the effective
action on the external gauge field, for the particular case of a linear wall.
\section{The effective action}\label{sec:effect}
\subsection{The model}\label{ssec:themodel}
We want to derive a general expression for the effective action due to a
massive Dirac field in the presence of an external Abelian gauge field, in a
spacetime region ${\mathcal U}$ with a non-trivial (static) spatial boundary
${\mathcal C}$. We shall assume ${\mathcal C}$ to correspond to a simple
closed plane curve ${\mathcal C}$ (figure 1).

The physical system, Dirac fermions in a background Abelian gauge field, may
be conveniently defined by its Euclidean action $S_f$ which, in our
conventions, is given by:
\begin{equation}\label{eq:defsf}
S_f({\bar\psi},\,\psi,\, A) \;=\; \int_{\mathcal U} \, d^3x \; {\bar\psi}(x)
\big(\not \!\! D  + m \big) \, \psi (x)
\end{equation}
where $\not \!\! D \equiv \gamma_\alpha D_\alpha$ and $D_\alpha \equiv
\partial_\alpha + i e A_\alpha(x)$, $\gamma_\alpha$ are Dirac's matrices and
$A_\alpha$ denotes an external Abelian gauge field. We shall adopt the
prescription that indices from the beginning of the Greek alphabet
$(\alpha,\,\beta,\ldots)$ can take the values $0$, $1$ and $2$, those from the
middle ($\mu,\,\nu,\ldots$) run from $0$ to $1$, while Roman indices
($i,\,j,\ldots$) can take the `spatial' values $1$ or $2$.  Dirac's matrices
are chosen according to the convention: \mbox{$\gamma_0 \equiv \sigma_1$},
\mbox{$\gamma_1 \equiv \sigma_2$} and \mbox{$\gamma_2 \equiv \sigma_3$}
($\sigma_1$, $\sigma_2$ and $\sigma_3$: Pauli's matrices) unless explicitly
stated otherwise.

In order to introduce the boundary conditions, we shall assume that the curve
${\mathcal C}$ has been parametrized: $\zeta \; \longrightarrow \; {\mathbf
  r}(\zeta)$, where ${\mathbf r}(\zeta) = \big(r_1(\zeta), \,
r_2(\zeta)\big)$, and that the parameter $\zeta$ belongs to some interval $I$.
Besides, for every point of ${\mathcal C}$, we introduce the unit vectors
${\mathbf{\hat t}}$ and ${\mathbf{\hat n}}$, tangent and (outer) normal to
${\mathcal C}$, respectively (see Figure 1).
\begin{figure}
\includegraphics[width=7cm]{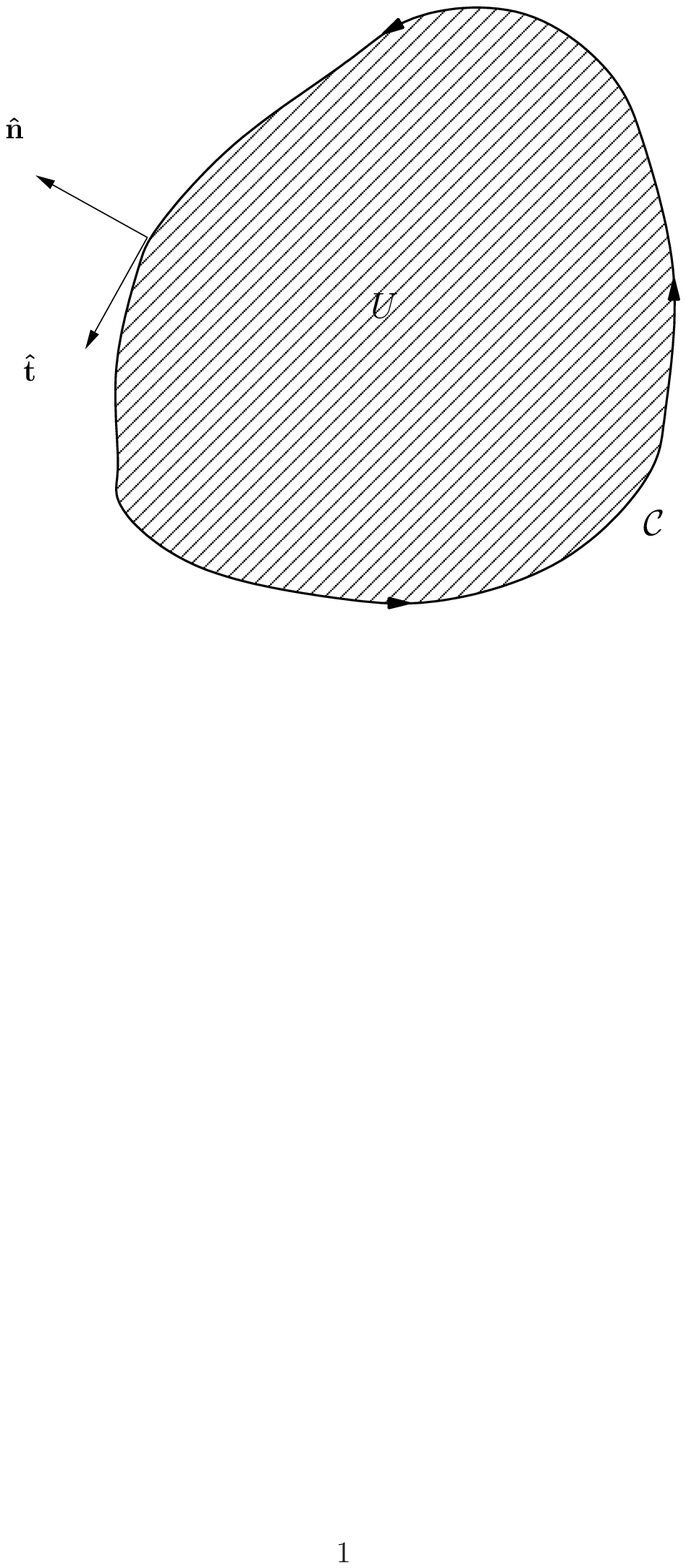}
\caption{The spatial region $U$, bounded by ${\mathcal C}$.
${\mathbf{\hat t}}$ and ${\mathbf{\hat n}}$ denote the unit
tangent and normal vectors, respectively.} \label{fig:region}
\end{figure}

An explicit expression for ${\mathbf{\hat t}}$ and ${\mathbf{\hat n}}$ may be
written as follows:
\begin{equation}\label{eq:deftn}
t_i (\zeta) \;=\;
\frac{\dot{r}_i(\zeta)}{|\dot{\mathbf{r}}(\zeta)|} \;,\;\;\;
n_i (\zeta) \;=\; \varepsilon_{ij} \, t_j (\zeta)\;,
\end{equation}
where $\dot{r}_i(\zeta) \equiv \frac{dr_i(\zeta)}{d\zeta}$.

Besides, when considering the large-mass limit, we shall also need to invoke
an alternative description for the curve ${\mathcal C}$, obtained by
introducing $u_1$ and $u_2$, two orthogonal curvilinear coordinates for the
plane, in such a way that ${\mathcal C}$ corresponds to $u_2 = 0$.  Since they
are orthogonal coordinates, the square of $d{\mathbf x}$ can be written as
follows:
\begin{equation}
|d{\mathbf x}|^2 \;=\; h_1^2\, du_1^2 \,+\, h_2^2 \, du_2^2 \;,
\end{equation}
where $h_1$ and $h_2$ may depend on $u_1$ and $u_2$.  A further simplification
we shall adopt is that we will fix $u_1$ to coincide with the arc length for
the points on the curve ${\mathcal C}$ (of course, when $u_2 = 0$), namely,
\begin{equation}
u_2 = 0 \;,\;du_2 = 0 \;\longrightarrow\;|d{\mathbf x}|^2 \;=\; du_1^2 \;=\; d\xi_1^2
\;.
\end{equation}
We shall not need to construct that system of coordinates
explicitly; rather, we note that, in a neighborhood of $u_2 = 0$,
one can construct $u_2$-constant coordinate lines by dragging
${\mathcal C}$ along the direction of ${\mathbf{\hat n}}$. On the
other hand, the $u_1$-constant lines are obtained by using the
property that ${\mathbf{\hat n}}$ is tangent to them (at every
point on the curve).

Equipped with the previous definitions, we introduce bag-like boundary
conditions on ${\mathcal C}$ for the fields $\psi$ and ${\bar\psi}$, as
follows:
\begin{equation}\label{eq:bagbc}
{\mathcal P}_L(\zeta) \, \psi (x_0,{\mathbf r}(\zeta)) \;=\; 0  \;,\;\;\;
{\bar \psi} (x_0,{\mathbf r}(\zeta)) \, {\mathcal P}_R (\zeta) \;=\; 0
\;\;,\;\;\; \forall \zeta \in I \;,
\end{equation}
where ${\mathcal P}_L$ and ${\mathcal P}_R$ are the projectors:
\begin{equation}
{\mathcal P}_L(\zeta)\;=\; \frac{1 \,+ \, {\mathbf \gamma} \cdot
{\mathbf{\hat n}}(\zeta)}{2}\;,\;\;
{\mathcal P}_R(\zeta)\;=\; \frac{1 \,- \, {\mathbf \gamma} \cdot
{\mathbf{\hat n}}(\zeta)}{2} \;,
\end{equation}
where the dot denotes the scalar product between (spatial) $2$-component
vectors: ${\mathbf a} \cdot {\mathbf b} \equiv a_i b_i = a_1 b_1 + a_2 b_2$.
The conditions (\ref{eq:bagbc}) ensure the vanishing, at all the points of
${\mathcal C}$, of $j_n$, the normal component of the induced fermion current:
\begin{equation}\label{eq:vnc}
j_n(x_0,{\mathbf r}(\zeta)) \;\equiv\; i e \Big\langle
{\bar\psi}\big( x_0,{\mathbf r}(\zeta) \big)
{\mathbf \gamma} \cdot {\mathbf{\hat n}}(\zeta) \psi
\big( x_0, {\mathbf r}(\zeta) \big) \Big\rangle \,=\, 0\;, \;\; \forall \zeta \in I\,.
\end{equation}
Here, the vacuum average $\langle \ldots \rangle$ is defined by:
\begin{equation}\label{eq:defvacav}
\langle \ldots \rangle \;\equiv\; \frac{\int_{\mathcal U} {\mathcal D}\psi \, {\mathcal
D}{\bar\psi} \ldots  e^{- S_f({\bar\psi},\psi; A)}}{\int_{\mathcal U}
{\mathcal D}\psi \, {\mathcal D}{\bar\psi}  e^{- S_f({\bar\psi},\psi; A)}}
\end{equation}
where $\int_{\mathcal U}$ means that the integration is constrained to verify
the proper boundary conditions, we shall see how to implement them by the use
of Lagrange multipliers (see below).

\subsection{Functional representation for the effective
action}\label{ssec:gamma}
Following the idea of the approach presented in~\cite{kardar}, we introduce
${\mathcal Z}(A)$, the partition function, and $\Gamma(A)$, the effective action
corresponding to the fluctuating Dirac field subject to the conditions
(\ref{eq:bagbc}), by means of the functional integral
\begin{eqnarray}\label{eqa:frepgamma}
{\mathcal Z}(A) &=& e^{ - \Gamma (A) } \;=\;
\int {\mathcal D} \psi {\mathcal D} {\bar\psi}\;
{\mathcal D} \chi_R {\mathcal D} {\bar\chi}_R
\; e^{- S_f ({\bar\psi},\psi, A) } \,
\nonumber\\
&\times& e^{ i \int dx_0 \int d\zeta \big[ {\bar \chi}_R(x_0,\zeta)
{\mathcal P}_L(\zeta) \, \psi (x_0,{\mathbf r}(\zeta))
+ {\bar \psi} (x_0,{\mathbf r}(\zeta)){\mathcal P}_R(\zeta) \chi_R (x_0,\zeta)
\big]}\;,
\end{eqnarray}
where we introduced auxiliary chiral Grasmmann fields $\chi_R$, ${\bar\chi}_R$
to exponentiate the $\delta$ functions. They are two-component fields
living in $1+1$ dimensions, and we find it convenient to use a more symmetric
notation for their arguments: $\chi_R = \chi_R(\xi_0,\xi_1)$,  ${\bar\chi}_R =
{\bar\chi}_R(\xi_0,\xi_1)$, where $\xi_0 \equiv x_0$ and $\xi_1 \equiv \zeta$.
These chiral fields may, of course, be thought of as chiral projections of Dirac
fiels:
\begin{eqnarray}
\chi_R(\xi_0,\xi_1) &=& {\mathcal P}_R(\xi_1) \chi(\xi_0,\xi_1) \nonumber\\
{\bar\chi_R}(\xi_0,\xi_1) &=& {\bar \chi}(\xi_0,\xi_1) {\mathcal
P}_L(\xi_1) \;.
\end{eqnarray}

We note that the auxiliary fields functional integration measure is:
\begin{equation}\label{eq:defintm}
{\mathcal D}\chi_R \, {\mathcal D}{\bar\chi}_R \;=\; \prod_{-\infty < \xi_0 <
\infty} \prod_{\xi_1 \in I} \Big[d\chi_R (\xi_0,\xi_1) \,
d{\bar\chi}_R(\xi_0,\xi_1) \Big] \;.
\end{equation}

We see in (\ref{eqa:frepgamma}) that the auxiliary fields will have a
non-trivial dynamics as a result of the Dirac field fluctuations.
Indeed, performing the (Gaussian) integral over the Dirac fields $\psi$,
${\bar\psi}$:
\begin{equation}\label{eq:zf1}
{\mathcal Z}(A) \;=\; \det \big( \not \!\! D + m \big) \;
\int \; {\mathcal D} \chi_R {\mathcal D} {\bar\chi}_R \;
e^{- \int d^2 \xi \int d^2\xi' {\bar{\chi}_R}(\xi)\,
{\mathcal K}_{\mathcal C}(\xi, \xi')\,\chi_R (\xi') }
\end{equation}
where we introduced:
\begin{equation}
{\mathcal K}_{\mathcal C} (\xi , \xi') \;=\;{\mathcal P}_L(\xi_1)\,
\langle \xi_0,{\mathbf r}(\xi_1) | \big(\not \!\! D + m\big)^{-1}
|{\xi'}_0,{\mathbf r}({\xi'}_1)\rangle {\mathcal P}_R(\xi'_1) \,,
\end{equation}
which is a kernel that induces a non-local action for the auxiliary fields.
Here, and for the rest of this article, we use a `Dirac bracket' notation
in order to simplify and clarify the formulae involving operator kernels.

Note that only one `chirality' of the auxiliary fields is actually coupled, but
the decomposition between the would-be `left' and `right' components is
point-dependent. This means, in particular, that
\mbox{$\bar{\chi}_R(\xi) \chi_R (\xi')$} does not necessarily vanish when
$\xi \neq \xi'$.
This fact prevents the introduction of one-component Weyl fermions as
auxiliary fields, since their local (point dependent) definitions would
render the apparent simplification illusory.  We shall however, in  some
special situations, use Weyl fermions: that will be the case when the
normal vector ${\hat n}$ is piecewise constant, like in the calculation of
the Casimir effect for parallel `plates' (lines).

The determinant factor on the rhs of (\ref{eq:zf1}) agrees with the would-be ${\mathcal
Z}(A)$ when the borders are sent to infinity (i.e., when there are no
borders). Since we are interested precisely in the effects due to the
presence of borders, we shall factor out that contribution, considering instead:
\begin{eqnarray}\label{eq:zft}
{\mathcal Z}_{\mathcal C}(A) &\equiv& \frac{{\mathcal Z}(A)}{\det \big(
\not \!\! D + m \big)} \;\equiv\; e^{-\Gamma_{\mathcal C}(A)}\nonumber\\
&=& \int \; {\mathcal D} \chi_R {\mathcal D} {\bar\chi}_R \;
e^{ - \int d^2 \xi \int d^2\xi' \bar{\chi}_R(\xi) {\mathcal K}_{\mathcal C}(\xi, \xi')
\chi_R (\xi') }\nonumber\\
&=& \det {\mathcal K}_{\mathcal C} \;.
\end{eqnarray}
Thus, the effective action corresponding to this functional is given by
\begin{equation}\label{eq:defgf}
\Gamma_{\mathcal C}(A) \;=\; - {\rm Tr} \, \ln {\mathcal K}_{\mathcal C} \;.
\end{equation}

At this point, it is useful to disentangle from $\Gamma_{\mathcal C}(A)$
the purely Casimir energy contribution from the part due to the external
field:
\begin{equation}\label{eq:defcas}
\Gamma_{\mathcal C}(A) \;=\;\Gamma_{\mathcal C}(0)
\,+\,{\widetilde{\Gamma}}_{\mathcal C}(A)
\end{equation}
where $\Gamma_{\mathcal C}(0)$ is proportional to the Casimir energy density
${\mathcal E}$, while ${\widetilde{\Gamma}}_{\mathcal C}(A)$, which
vanishes when $A=0$, is a measure of the effect of the borders on the
response of the system to the external field.

We shall use a ${\mathcal Z}$ functional corresponding to each of these terms;
they will be denoted by ${\mathcal Z}_{\mathcal C}(0)$ and
$\widetilde{\mathcal Z}_{\mathcal C}(A)$ (in an obvious notation).
\subsection{Gauge invariance of $\Gamma(A)$}\label{sec:ginv}
Being a functional of $A$, the study of gauge invariance for $\Gamma(A)$,
reduces to an analysis of its behaviour under gauge transformations for the
gauge field, namely:
\begin{equation}
\delta_\omega \Gamma(A) \;=\; \Gamma(A + \partial \omega) \,-\, \Gamma(A)
\end{equation}
where $\omega$ is a smooth function of (all of) the spacetime coordinates.  In
order to understand the effect of those transformations, it is convenient to
recall representation (\ref{eqa:frepgamma}), in order to see that:
\begin{eqnarray}
e^{ - \Gamma (A\,+\,\partial \omega) } &=&
\int {\mathcal D} \psi {\mathcal D} {\bar\psi}\;
{\mathcal D} \chi_R {\mathcal D} {\bar\chi}_R
\; e^{- S_f ({\bar\psi},\psi, A + \partial \omega) } \nonumber\\
&\times & \exp\Big\{ i \int d^2\xi \big[ {\bar \chi}(\xi)
{\mathcal P}_L(\xi_1) \, \psi (\xi_0,{\mathbf r}(\xi_1))\nonumber\\
&+& {\bar \psi} (\xi_0,{\mathbf r}(\xi_1)){\mathcal P}_R(\xi_1)
\chi (\xi) \big]\Big\}\; .
\end{eqnarray}
We then compensate the change in $S_f$ due to the transformation of $A$, by
means of a gauge transformation in the Dirac fields:
\begin{equation}\label{eq:ward2}
\psi(x) \;\to\; e^{-i e \omega(x)} \, \psi(x) \;\;,\;\;\;\;
{\bar\psi}(x) \;\to\;  {\bar\psi}(x)\, e^{i e \omega(x)} \;,
\end{equation}
which is, of course, non anomalous. The only source of non-invariance under the
transformations we have just performed is in the coupling to the Lagrange
multiplier fields, which is concentrated on the boundary ${\mathcal C}$:
\begin{eqnarray}\label{eqa:ward3}
e^{- \Gamma (A\,+\,\partial \omega) } &=& \int {\mathcal D} \psi {\mathcal D} {\bar\psi}\;
{\mathcal D} \chi_R {\mathcal D} {\bar\chi}_R \; e^{- S_f ({\bar\psi},\psi, A) }
\nonumber\\
&\times& \exp\Big\{ i \int d^2\xi \big[ {\bar \chi}_R(\xi)
{\mathcal P}_L(\xi_1) \, e^{-i e \omega(\xi_0,{\mathbf r}(\xi_1))}\,
\psi (\xi_0,{\mathbf r}(\xi_1))
\nonumber\\
&+& {\bar \psi} (\xi_0,{\mathbf r}(\xi_1)) \, e^{i e \omega(\xi_0,{\mathbf r}(\xi_1))}\,
{\mathcal P}_R(\xi_1) \chi_R (\xi) \big] \Big\}\;.
\end{eqnarray}
At this point, we realise that all the dependence in $\omega$ can be erased by
transforming the Lagrange multipliers:
\begin{equation}\label{eq:ward4}
\chi_R(\xi) \;\to\; e^{-i e \omega(\xi_0,{\mathbf r}(\xi_1))} \, \chi_R(\xi)
\;\;,\;\;\;\;
{\bar\chi}_R(\xi) \;\to\; {\bar\chi}_R(\xi)\, e^{i e \omega(\xi_0,{\mathbf
r}(\xi_1))} \;.
\end{equation}
Since they are chiral fields, there arises a non-trivial Jacobian ${\mathcal
J}(\omega,A)$ from their integration measure:
\begin{equation}\label{eq:defjac}
{\mathcal D} \chi_R {\mathcal D} {\bar\chi}_R \;\to \;
{\mathcal D} \chi_R {\mathcal D} {\bar\chi}_R  \;{\mathcal J}(\omega, A) \;.
\end{equation}
To the first order in $\omega$
\begin{equation}\label{eq:formjac}
{\mathcal J}(\omega, A) \;\simeq\;\exp \big[i e \int d^2\xi \,
\omega(\xi_0,{\mathbf r}(\xi_1)) \; {\mathcal F}(A;\xi_0,\xi_1)\big]
\end{equation}
where ${\mathcal F}(A;\xi_0,\xi_1)$ is the anomaly a functional of $A$ and a
function of the parameters of the worldsheet corresponding to the border. We
have assumed that $\xi_1 \equiv u_1$, so that the ${\mathcal C}$ coincides with
$u_2=0$.

From (\ref{eqa:ward3}), (\ref{eq:defjac}) and (\ref{eq:formjac}) we
conclude that:
\begin{equation}\label{eq:ward5}
\partial_\mu \Big[\frac{\delta \Gamma(A)}{\delta A_\mu(x)}\Big]\;=\;
i \, e \, \int d\xi_1 \, \delta({\mathbf x} - {\mathbf r}(\xi_1)) \;
{\mathcal F}(A;x_0,\xi_1) \;,
\end{equation}
which shows explicitly the fact that the gauge non-invariance will be
concentrated on the boundary, although the actual form of the anomaly will, in
principle, depend on the field $A$ also at points slightly away from the
boundary.

We see that (\ref{eq:ward5}) is relevant to the physical problem
of imposing bag-like boundary conditions. Indeed, we easily see
that (\ref{eq:ward5}) implies:
\begin{equation}\label{eq:ward6}
\partial_\alpha j_\alpha(x)\;=\;-
i \, e \, \int d\xi_1 \, \delta({\mathbf x} - {\mathbf r}(\xi_1)) \;
{\mathcal F}(A;x_0,\xi_1) \;,
\end{equation}
where $j_\alpha(x)$ is the induced vacuum current:
\begin{equation}\label{eq:vtc}
j_\alpha(x) \;\equiv\; i e \big\langle {\bar\psi}(x)\gamma_\alpha \psi (x)\big\rangle \;.
\end{equation}
Integrating the anomalous divergence equation (\ref{eq:ward6}) on the
world-volume generated by the (fixed) region ${\mathcal U}$ during a time
interval $[0,T]$, we see that Gauss' theorem yields:
$$
\int_{\mathcal U} d{\mathbf x} \, j_0(0,{\mathbf x}) \,-\, \int_{\mathcal U}
d{\mathbf x} \, j_0(T,{\mathbf x})
\;=\; \int_{{\mathcal C}\times [0,T]} \,d^2\xi \,j_n(x_0,{\mathbf
r}(\zeta))
$$
\begin{equation}
+ i \, e \, \int d^2\xi \, {\mathcal F}(A;\xi_0,\xi_1) \;.
\end{equation}
Then the existence of the anomaly implies that, under some circumstances, the
bag condition will be violated. Indeed, assuming for example that the total
charge of the $2+1$ dimensional system is constant (insulated system), then the
lhs of the previous equation vanishes, and we get a relation involving the
integral of the anomaly and the flux of the current. If the former is not zero,
the latter is necessarily different from zero. The explicit form for the
anomaly is, in these coordinates ($\xi_1 \equiv u_1$, $\xi_2 \equiv u_2$):
\begin{equation}
{\mathcal F}(A;\xi_0,\xi_1) \;=\; - \frac{e}{2\pi} \, \varepsilon_{\mu\nu}
\partial_\mu {\tilde A}_\nu(\xi) \;,
\end{equation}
where ${\tilde A}_\mu = A_\mu (\xi_0,\xi_1,0)$. Thus the non-vanishing of the
anomlous contribution depends only on the circulation of ${\tilde A}_1$
(which is the tangential component of $A$ on ${\mathcal C}$) at the times $T$
and $0$. This may also be put in terms of the {\em magnetic} flux through
${\mathcal U}$ at those times. Then:
\begin{eqnarray}
 \int_{{\mathcal C}\times [0,T]} \,d^2\xi \,j_n(x_0,{\mathbf
r}(\zeta)) &=& \frac{e^2}{2\pi} \Big[ \int d{\mathbf x}
\varepsilon_{ij}\partial_i A_j ({\mathbf x},T) \nonumber\\
&-& \int d{\mathbf x} \varepsilon_{ij}\partial_i
A_j ({\mathbf x},0) \Big] \;.
\end{eqnarray}
This anomalous current flux is of course just another manifestation of the fact
that the effective theory shall contain a Chern-Simons like term, which
introduces a gauge non-invariance on the boundary. Indeed, that is the
usual set-up for the study of this phenomenon, which is dealt with in the
context of the effective theory for the bulk, and the dynamics for the boundary
modes is obtained therefrom~\cite{wen}.

Of course, the gauge non-invariance could be cured by adjusting the matter
content, or by imposing conditions on the external gauge field, like the
invariance of the total magnetic flux through ${\mathcal U}$.

\subsection{Fermion propagator}\label{ssec:fprop}
Let us derive now an expression for the fermion propagator by using this
representation. A simple way to do that is to introduce a generating
functional containing linear couplings to two auxiliary Grassmann source,
denoted by ${\bar\eta}$ and $\eta$:
\begin{eqnarray}
{\mathcal Z}({\bar\eta},\eta)
&=& \int {\mathcal D} \psi {\mathcal D} {\bar\psi}\;
{\mathcal D} \chi {\mathcal D} {\bar\chi}
\; e^{- S_f ({\bar\psi},\psi, 0) + \int d^3x \, ({\bar\eta}\psi +
{\bar\psi}\eta) } \nonumber\\
&\times& e^{ i \int dx_0 \int d\xi \big[ {\bar \chi}(x_0,\xi)
{\mathcal P}_L(\xi) \, \psi (x_0,{\mathbf r}(\xi))
+ {\bar \psi} (x_0,{\mathbf r}(\xi)){\mathcal P}_R(\xi) \chi (x_0,\xi)
\big]} \;,
\end{eqnarray}
whereby the fermion propagator $\langle \psi(x) {\bar\psi}(y)\rangle$ can
be obtained as follows:
\begin{equation}
\langle \psi(x) {\bar\psi}(y)\rangle \;=\; \frac{1}{{\mathcal Z}(0,0)}
\,\Big[\frac{\delta^2}{\delta\eta(y) \,\delta{\bar\eta}(x)} {\mathcal
Z}({\bar\eta},\eta) \Big]\Big|_{\eta=0,{\bar\eta=0}} \;.
\end{equation}
Performing the Gaussian integrations, and evaluating the derivatives, we
obtain for the free fermion propagator the following expression:
$$
\langle \psi(x) {\bar\psi}(y)\rangle \;=\;
\langle x|(\not\!\partial + m)^{-1}| y\rangle \;-\;
\int d^2\xi' \int d^2\xi''
$$
\begin{equation}
\langle x| (\not\!\partial + m)^{-1}| \xi'_0,{\mathbf r}(\xi'_1)\rangle
{\mathcal P}_R(\xi'_1)
{\mathcal K}_{\mathcal C}^{-1}(\xi',\xi'') {\mathcal P}_L(\xi''_1)
 \langle \xi''_0,{\mathbf r}(\xi''_1) | (\not\!\partial + m)^{-1}| y \rangle \;.
\label{freeprop}
\end{equation}
It is evident, from the previous expression, that the propagator so obtained
does verify the proper boundary conditions. Indeed,
$$
{\mathcal P}_L(\xi_1) \langle \psi\big(\xi_0,{\mathbf r}(\xi_1)\big) \, {\bar\psi}(y)\rangle \,=
\,{\mathcal P}_L(\xi_1) \langle \xi_0,{\mathbf r}(\xi_1)| (\not\!\partial +
m)^{-1}| y\rangle \;
$$
$$
-\; \int d^2\xi' \int d^2\xi'' \, {\mathcal P}_L(\xi_1)
\langle \xi_0,{\mathbf r}(\xi_1) | (\not\!\partial + m)^{-1}|
\xi'_0,{\mathbf r}(\xi'_1)\rangle
$$
$$
{\mathcal P}_R(\xi'_1) {\mathcal K}_{\mathcal
C}^{-1}(\xi',\xi'') {\mathcal P}_L(\xi''_1)
 \langle \xi''_0,{\mathbf r}(\xi''_1) | (\not\!\partial + m)^{-1}| y \rangle
$$
$$
=\; {\mathcal P}_L(\xi_1)\langle \xi_0,{\mathbf r}(\xi_1)| (\not\!\partial
+ m)^{-1}| y\rangle -\; \int d^2\xi' \int d^2\xi'' \;  {\mathcal K}_{\mathcal C}(\xi,\xi')
{\mathcal K}_{\mathcal C}^{-1}(\xi',\xi'')
$$
$$
{\mathcal P}_L(\xi''_1) \langle \xi''_0,{\mathbf r}(\xi''_1) | (\not\!\partial
+ m)^{-1}| y \rangle \;= \;{\mathcal P}_L(\xi_1)  \langle \xi_0,{\mathbf r}(\xi_1)| (\not\!\partial
+ m)^{-1}| y\rangle
$$
\begin{equation}
-\;\int d^2\xi'' \; \delta(\xi_0 - \xi''_0) \,\delta(\xi_1-\xi''_1)
\; {\mathcal P}_L(\xi''_1) \, \langle \xi''_0,{\mathbf r}(\xi''_1)| (\not\!\partial + m)^{-1}| y\rangle
\;=\; 0\;.
\end{equation}

In section~\ref{sec:linear}, we will find an explicit expression
for the free fermion propagator in the presence of a linear wall.

\section{Casimir energy}\label{sec:casimir}
Let us consider here the Casimir term $\Gamma_{\mathcal C}(0)$,
for the physically interesting case of $m=0$, evaluating it
explicitly for a particular geometry.

We first write this object more explicitly, in terms of the corresponding
functional integral over auxiliary fields:
\begin{equation}
{\mathcal Z}_{\mathcal C} (0)=
e^{-\Gamma_{\mathcal C}(0)} =\int {\mathcal D} \chi {\mathcal D} {\bar\chi}
e^{-\int d^2 \xi \int d^2\xi' \bar{\chi}_R(\xi) {\mathcal P}_L(\xi_1)
\langle \xi_0, {\mathbf r}(\xi_1) | \not\partial^{-1} | \xi_0, {\mathbf r}(\xi_1')
\rangle {\mathcal P}_R(\xi_1')\chi_R (\xi') }\,.
\end{equation}
The simplest non-trivial geometry is the one corresponding to the
region: ${\mathcal U} = \{ (x_1,x_2): 0 \leq x_2 \leq l\}$, so
that ${\mathcal
  C}$ is just the union of two lines: ${\mathcal C}_0$, corresponding to
$x_2=0$ and ${\mathcal C}_l$, to $x_2=l$.  Of course, in this case, the normal
vectors shall be $-{\mathbf{\hat x}_2}$ and ${\mathbf{\hat x}_2}$,
respectively. In order to parametrize the auxiliary fields, we find it
convenient to use $x_1 \in (-\infty, +\infty)$ as the (common) parameter, but
using a label to distinguish the fields corresponding to the lower
($\chi^{(0)}(x_0,x_1)$), and upper ($\chi^{(l)}(x_0,x_1)$) borders.

Then,
\begin{equation}
e^{- \Gamma_{\mathcal C}(0)} \;=\; \int \; {\mathcal D} \chi^{(0)} {\mathcal
D} {\bar\chi}^{(0)} \; {\mathcal D} \chi^{(l)} {\mathcal D}
{\bar\chi}^{(l)} \; e^{- S_{\mathcal C}( \chi^{(0)},
{\bar\chi}^{(0)}\, ;\; \chi^{(l)} ,  {\bar\chi}^{(l)})}
\end{equation}
where the `action' $S_{\mathcal C}$ is defined by:
\begin{eqnarray}
S_{\mathcal C} &=&\int d^2x  \int d^2x' \Big[
\bar{\chi}^{(0)}(x) {\mathcal P}_-
\langle x_0,x_1,0 | \not\!\partial^{-1} | x_0', x_1',0\rangle {\mathcal P}_+
\chi^{(0)} (x')  \nonumber\\
&+&\bar{\chi}^{(l)}(x) {\mathcal P}_+
\langle x_0,x_1,l | \not\!\partial^{-1} | x_0', x_1',l\rangle {\mathcal P}_-
\chi^{(l)} (x')  \nonumber\\
&+& \bar{\chi}^{(0)}(x) {\mathcal P}_-
\langle x_0,x_1,0 | \not\!\partial^{-1} | x_0', x_1',l\rangle {\mathcal P}_-
\chi^{(l)} (x')  \nonumber\\
&+& \bar{\chi}^{(l)}(x) {\mathcal P}_+
\langle x_0,x_1,l | \not\!\partial^{-1} | x_0', x_1',0\rangle {\mathcal P}_+
\chi^{(0)} (x') \Big]
\end{eqnarray}
where ${\mathcal P}_\pm = \frac{1 \pm \gamma_2}{2}$. It should be clear now
that, since these projectors are constant, the auxiliary fields
$\chi^{(0,l)}$, when multiplied by those projectors, are trivial functions of
(different) one-component Weyl fermions.  Namely,
\begin{equation}
{\mathcal P}_+ \, \chi^{(0)} (x) \;=\; \left( \begin{array}{c} \eta^{(0)} (x)\\ 0
\end{array} \right) \;,\;\;
{\mathcal P}_- \, \chi^{(l)} (x) \;=\; \left( \begin{array}{c} 0 \\
\eta^{(l)}(x)
\end{array} \right) \;,
\end{equation}
and
\begin{equation}
{\bar \chi}^{(0)}(x)\, {\mathcal P}_- \;=\; ( 0 \,,\;{\bar\eta}^{(0)} (x) )
\;,\;\;
{\bar \chi}^{(l)}(x) \, {\mathcal P}_+ \;=\; ({\bar\eta}^{(l)} (x) \,,\; 0 ) \;,
\end{equation}
where $\eta^{(0)}$, $\eta^{(l)}$, and their adjoints, are one-component Weyl
fields.

We may combine them into a two-component field $\chi$:
\begin{equation}
\chi(x) \;\equiv \; \left( \begin{array}{c} \eta^{(0)} (x)\\ \eta^{(l)}(x)
\end{array} \right) \;, \;\;
{\bar\chi}(x) \;\equiv\; ( {\bar\eta}^{(0)} (x) \;,\; {\bar\eta}^{(l)}(x)) \;,
\end{equation}
and write the action $S_{\mathcal C}$ as:
\begin{equation}
S_{\mathcal C} \;=\;\int d^2x  \int d^2x' \; {\bar\chi}(x) {\mathcal
D}(x,x') \chi (x') \;,
\end{equation}
where
\begin{equation}\label{eq:defm}
{\mathcal D}(x,x') \;=\; \left(
\begin{array}{cc}
\langle x_0,x_1,0 | \frac{\partial^+}{\partial^2} | x_0',
x_1',0\rangle &
\langle x_0,x_1,0 | -\frac{\partial_2}{\partial^2} | x_0',
x_1',l\rangle \\
\langle x_0,x_1,l | \frac{\partial_2}{\partial^2} | x_0', x_1',0\rangle  &
\langle x_0,x_1,l | \frac{\partial^-}{\partial^2} | x_0', x_1',l\rangle
\end{array}
\right) \;,
\end{equation}
where $\partial^+ \equiv \partial_0 + i \partial_1$ and $\partial^- \equiv
\partial_0 - i \partial_1$.
Then we have,
\begin{equation}\label{eq:loga}
\Gamma_{\mathcal C}(0) \;=\; - {\rm Tr} \ln {\mathcal D} \;=\; - \frac{1}{2}
  {\rm Tr} \ln \big({\mathcal D}^\dagger {\mathcal D}\big)\;,
\end{equation}
which is best evaluated by introducing a Fourier transformation with respect
to the coordinates $x_0$ and $x_1$.  We see that:
\begin{equation}\label{eq:defmk}
\widetilde{\mathcal D}(k) \;=\; \left(
\begin{array}{cc}
-i (k_0 + i k_1)/2 k
& e^{- l k}/2 \\
e^{- l k}/2
&
-i (k_0 - i k_1)/2 k
\end{array}
\right) \;.
\end{equation}
Then:
\begin{equation}
\Gamma_{\mathcal C}(0) \;=\; - \frac{1}{2} \, L T \, \int \frac{d^2k}{(2\pi)^2} \;
\ln \big[ \frac{1}{2} \, ( 1 + e^{- 2 l k})  \big] \;,
\end{equation}
where $L$ is the length of the plates, and $T$ the extension of the
(Euclidean) time interval. In this expression, there is a (divergent)
$l$-independent contribution which we attribute to the self-energy or each
plate, plus a Casimir energy (energy per unit length):
\begin{equation}
{\mathcal E} \;=\; - \frac{1}{2} \, \int \frac{d^2k}{(2\pi)^2} \;
\ln \big( 1 + e^{- 2 l k} \big) \;, \label{intcas}
\end{equation}
which can be easily integrated:
\begin{equation}
{\mathcal E} \;=\; - \frac{3 \zeta(3)}{64 \pi l^2} \,.
\end{equation}

An interesting feature of this result is that the Casimir energy is already
given by an {\em integral\/} over the momenta which are parallel to the plates.
Thus, the series over the discrete momenta along the normal direction to the
plates has already been summed up.

Of course, both approaches are related, as can be easily seen by first noting
that the eigenvalues of ${\mathcal D}^\dagger {\mathcal D}$ are identical to
the squares of the eigenvalues of a Dirac {\em Hamiltonian\/} in $3+1$
dimensions (with one of the spatial coordinates playing the role of the time).
Those eigenvalues are known to be~\cite{Chodos:1974dm}:
\begin{equation}\label{eq:eigen1}
\lambda_{n,k} \;=\; \sqrt{\omega_n^2 \,+\, k^2} \;,
\end{equation}
where $\omega_n = \frac{(2 n + 1) \pi}{2 l}$.  Then we see that:
\begin{equation}\label{eq:series}
\ln \big( 1 + e^{- 2 l k} \big) \;=\; \frac{1}{2} \sum_{n=-\infty}^\infty
\; \ln \Big[(2 l)^2 ( \omega_n^2 + k^2 ) \Big] \;,
\end{equation}
where we have neglected $l$-independent terms.  The sum on the rhs
of (\ref{eq:series}) arises naturally when one evaluates the
Casimir energy by finding the eigenvalues of the Dirac operator
for the modes constrained to satisfy the bag boundary conditions.

One can also obtain an expression similar to (\ref{intcas}) for
the Casimir energy starting from its usual definition as the sum
of the zero-point energies of the field modes, and using Cauchy's
theorem to write the sum as a contour integral in the complex
plane \cite{Milonni94}.
\section{The large-mass limit}\label{sec:largem}
We shall approach this limit by following two different strategies: first, we
shall begin with a quantized Dirac field, implementing the approximations and
simplifications that follow from the assumption that the fermionic mass is
much larger than the other relevant dimensionful objects; i.e., the gauge
field derivatives.  Our second approach amounts to start from the effective
`bosonized' theory that follows by taking the large-mass limit beforehand, and
introducing the boundary condition afterwards.
\subsection{Fermionic representation} In the large mass limit, we can
obtain some explicit results as a consequence of the fact that the kernel
${\mathcal K}_{\mathcal C}$ becomes {\em local}.  We begin by noting that
$\widetilde{\mathcal Z}_{\mathcal C}(A)$ may be regarded as a regularized
version (with the fermion mass $m$ playing the role of an $UV$ cutoff) of the
determinant of a {\em local\/} operator. Indeed, taking into account the
fact that ${\mathcal P}_L$ and ${\mathcal P}_R$ are orthogonal projectors
at every point of ${\mathcal C}$, we may rewrite ${\mathcal K}_{\mathcal C}$ as:
\begin{equation}
{\mathcal K}_{\mathcal C} (\xi , \xi') \;=\;
{\mathcal P}_L(\xi_1) \,
\langle \xi_0,{\mathbf r}(\xi_1)|
\frac{-\not \!\! D}{-\not \!\! D^2 + m^2}
|{\xi'}_0,{\mathbf r}({\xi'}_1)\rangle \,{\mathcal P}_R(\xi'_1)
\end{equation}
or:
\begin{equation}
{\mathcal K}_{\mathcal C} (\xi , \xi') \;=\; \frac{1}{m^2}\,
{\mathcal P}_L(\xi_1) \,
\langle \xi_0,{\mathbf r}(\xi_1)|
f\big(\frac{-\not \!\! D^2}{m^2}\big) \, \big(-\not \!\! D\big)
|{\xi'}_0,{\mathbf r}({\xi'}_1)\rangle \, {\mathcal P}_R(\xi'_1)
\end{equation}
where $f(x) \equiv \frac{1}{1 + x}$. Since $f(0)=1$, and $f$ and all its
derivatives tend to zero when $x \to \infty$, it is evident that
$\widetilde{\mathcal Z}_{\mathcal C}(A)$ is a regularized
version of another functional, which we denote by ${\mathcal Z}_{loc}(A)$,
defined as the result of taking the $f \to 1$ limit in
$\widetilde{\mathcal Z}_{\mathcal C}(A)$:
\begin{equation}\label{eq:zloc}
{\mathcal Z}_{loc} (A) \;=\;
\Big[ \widetilde{\mathcal Z}_{\mathcal C} (A)\Big]_{m \to \infty} \;=\;
\int \; {\mathcal D} \chi {\mathcal D} {\bar\chi} \;
e^{- S_{loc} \big({\bar\chi},\,\chi,\,A \big)} \;,
\end{equation}
where
\begin{equation}
S_{loc}\;=\;  \int d^2 \xi \;{\bar\chi}(\xi) \, {\mathcal K}_{loc}
(\xi,\xi') \;\chi (\xi) \;,
\end{equation}
and:
\begin{equation}\label{eq:defkloc}
 {\mathcal K}_{loc} (\xi,\xi') \,=\, - \, {\mathcal P}_L(\xi_1) \,
\langle \xi_0,{\mathbf r}(\xi_1)|\,\not \!\! D
|{\xi'}_0,{\mathbf r}({\xi'}_1)\rangle \, {\mathcal P}_R(\xi'_1) \;.
\end{equation}
$S_{loc}$ is a local action, and we have neglected an infinite
($A$-independent) factor $\det (m^{-2})$.  It is important at this point to
remark that, since the auxiliary fields behave as $1+1$ dimensional Dirac
fermions with a minimal gauge coupling, no infinity arises when removing
the regulator ($m \to \infty$). Of course, this will not necessarily be the
case in higher dimensions. Besides, the regulator only affects the real
part of the effective action (namely, the modulus of the fermionic
determinant). The imaginary part is of course still there, and requires its
own regularization. Note, however, that the imaginary part is also
determined by the local action, since the `regulator' affects only the
modulus of the eigenvalues of the Dirac operator, and those are gauge
invariant.

It should be obvious that, to make further progress, it is convenient to
write (\ref{eq:defkloc}) more explicitly, in terms of coordinates which are
more adapted to the geometry of ${\mathcal C}$. To that end, we invoke the
coordinates $u_1$ and $u_2$, introduced in the previous section, recalling
that $u_1$ and $\xi_1$ actually coincide on ${\mathcal C}$, to see
that the local action may be written as follows:
\begin{equation}
S_{loc} \;=\; \int d^2\xi \; {\bar\chi}(\xi_0,\xi_1) \,
{\mathcal P}_L(\xi_1) ( \tilde{\gamma}_\mu d_\mu ) {\mathcal P}_R(\xi_1) \,
\chi(\xi_0,\xi_1)
\end{equation}
where:
$$
{\mathcal P}_L (\xi_1) \;=\; \frac{1 \,+\, {\tilde \gamma}_2(\xi_1)}{2}
\;,\;\;\;{\mathcal P}_R (\xi_1) \;=\; \frac{1 \,-\, {\tilde \gamma}_2(\xi_1)}{2}
$$
$$
d_\mu \;\equiv \; \partial_\mu \,+\, i e {\tilde A}_\mu(\xi_0,\xi_1)
$$
\begin{equation}
{\tilde A}_\mu (\xi_0,\xi_1) \;\equiv\; A_\mu \big(\xi_0, r_1(\xi_1),
r_2(\xi_1)\big)
\end{equation}
and
\begin{equation}
\tilde{\gamma}_0 \,=\, \gamma_0 \;,\;\;\;
\tilde{\gamma}_1 (\xi_1) \,=\, {\mathbf \gamma} \cdot {\mathbf{\hat{t}}}(\xi_1)
\;,\;\;\;
\tilde{\gamma}_2 (\xi_1) \,=\, {\mathbf \gamma} \cdot {\mathbf{\hat{n}}}(\xi_1) \;.
\end{equation}

Note that there is no coupling to the component of $A$ that is normal to
the curve.

Then,  in the infinite mass limit, the effective action due to the presence
of the boundary reduces to the one of a chiral fermion determinant:
\begin{equation}
{\mathcal Z}_{loc} (A) \;=\; e^{- \Gamma_{loc} (A)} \;,\;\;\;
\Gamma_{loc}(A) \;=\; - {\rm Tr} \ln \big[ \tilde{\gamma}_\mu d_\mu  {\mathcal
P}_R\big] \;.
\end{equation}

By our comment above on the imaginary part, it is clear that:
\begin{equation}\label{eq:imag1}
{\rm Im} \Gamma_{\mathcal C} (A) \;=\;
{\rm Im} \Gamma_{loc} (A) \;,
\end{equation}
where, of course, a regularization procedure has to be invoked (as it has to be
also in a local theory).
\subsection{Bosonic representation}
To describe a Dirac field coupled to an external gauge field $A_\alpha$ we
may, in the limit when the fermion mass is large (in comparison with the
momenta of the external fields) use an approximate bosonization
procedure~\cite{b1,b2,b3,b4}. The fermion $\leftrightarrow$ boson mapping
leads to a bosonic action, $S^{(b)}$, whose leading form in a large-mass
expansion is given by:
\begin{equation}\label{eq:bosact}
S^{(b)}(a,A) \;=\; S_{CS}(a) \,+\, i \, \int
d^3 x \,\varepsilon_{\alpha\beta\gamma}\, \partial_\beta a_\gamma A_\alpha
\;,
\end{equation}
where $a_\alpha$ is a new gauge field, introduced to implement the duality,
and $S_{CS}$ is the Chern-Simons action:
\begin{equation}
S_{CS}(a) \;=\; i \,\frac{\kappa}{2} \, \int d^3x \, \varepsilon_{\alpha\beta\gamma} a_\alpha \,\partial_\beta a_\gamma \;,
\end{equation}
where $\kappa$ is a constant.

The second term in (\ref{eq:bosact}) corresponds to the standard coupling
between current and external gauge field, since $a_\alpha$ is related to
the average value of the fermionic current, $j_\alpha$, by:
\begin{equation}\label{eq:dual}
j_\alpha \;=\; i \, \varepsilon_{\alpha\beta\gamma} \,\partial_\beta a_\gamma \;,
\end{equation}
a relation which is exact, i.e, independent of the approximation used to
obtain the bosonized action.

The bosonized partition function {\em in the absence of boundaries},
${\mathcal Z}^{(b)}(A)$, can be defined as follows:
\begin{equation}
{\mathcal Z}^{(b)}\;=\;\int {\mathcal D}a \, e^{-S^{(b)}(a,A)} \;.
\end{equation}

To take into account the boundary conditions corresponding to the fermionic
theory in this setting, we note that the mapping between the fermionic and
bosonic representations for the current implies that we should impose:
\begin{equation}
(\partial_0 a_l - \partial_l a_0) t_l  \,=\, 0 \;\;\; {\rm on} \; {\mathcal C}
\end{equation}
i.e., the wall must behave like a `perfect conductor' for the $a_\alpha$ gauge field,
since there is no tangential component for its electric field on the border.
Note that this boundary condition is independent of the large mass
expansion, since it only relies upon the exact mapping (\ref{eq:dual}) between
$j_\alpha$ and $a_\alpha$.

Introducing now a new field $\varphi(\xi_0,\xi_1)$, a Lagrange multiplier
field for the previous condition, we are lead to $\widetilde{\mathcal Z}_{\mathcal
C}^{(b)}$, the bosonized form of the partition function for the contribution due
to the modes localized on the borders:
\begin{equation}
{\widetilde{\mathcal Z}}^{(b)}_{\mathcal C}(A) \;=\; e^{- {\tilde
\Gamma}^{(b)}_{\mathcal C}(A)} \;=\;
\frac{{\mathcal Z}^{(b)}_{\mathcal C}(A)}{{\mathcal Z}^{(b)}_{\mathcal C}(0)}
\end{equation}
where now
\begin{eqnarray}
{\mathcal Z}^{(b)}_{\mathcal C}(A) &=& \frac{1}{{\mathcal Z}^{(b)} (A)} \;
\int {\mathcal D}a \,{\mathcal D}\varphi \,
e^{-S_{CS}(a) -i \int d^3x \varepsilon_{\alpha\beta\gamma} A_\alpha \partial_\beta
a_\gamma} \nonumber\\
&\times& e^{i\int d^2\xi \varphi(\xi_0,\xi_1) f_{0l}(\xi_0,{\mathbf r}(\xi_1))
t_l(\xi_1)}
\end{eqnarray}
where $f_{\alpha\beta} \equiv \partial_\alpha a_\beta - \partial_\beta a_\alpha$.

When evaluating the Gaussian integral, an important point arises as a
consequence of the existence of a boundary term coming from an integration by
parts. Indeed, to perform the Gaussian integral we need to rewrite the term
that couples the bosonized current to the external field $A_\alpha$.
After performing an integration by parts and applying Gauss' theorem, we see
that:
\begin{equation}
\int d^3 x
\,\varepsilon_{\alpha\beta\gamma}\, \partial_\beta a_\gamma A_\alpha
\;=\; \int d^3 x\,\varepsilon_{\alpha\beta\gamma}\, \partial_\beta A_\gamma
a_\alpha \,+\, \int d^3x \, a_\alpha(x) R_\alpha(x)
\end{equation}
where
\begin{eqnarray}
R_0 (x) &=& - \int d\xi_1 \, \delta({\mathbf x}-{\mathbf
r}(\xi_1)) \, A_l(x) \, t_l(\xi_1) \nonumber\\
R_k (x) &=& \int d\xi_1 \, \delta({\mathbf x}-{\mathbf
r}(\xi_1)) \, A_0(x) \, t_k(\xi_1) \;.
\end{eqnarray}
To keep this boundary term amounts to reproducing the proper result, in
particular for the anomalous behaviour of the effective action
under gauge transformations.

The Gaussian integral over $a_\alpha$ can now be performed, what yields an action
$S_{\mathcal C}$ for the Lagrange multiplier field:
\begin{equation}
{\mathcal Z}^{(b)}_{\mathcal C}(A) \;=\;\int {\mathcal D} \varphi
\; e^{-S_{\mathcal C} (\varphi, A)} \;,
\end{equation}
where a `bulk' Chern-Simons term has been cancelled out, and:
\begin{eqnarray}\label{eq:sc}
S_{\mathcal C} (\varphi, A) &=& \frac{1}{2} \, \int d^2 \xi \int d^2 \xi'
\, \big( \partial_0\varphi(\xi)-A_0(\xi_0,{\mathbf r}(\xi_1)\big) \,
t_j(\xi_1) M_{jk}(\xi,\xi') t_k(\xi_1')\, \nonumber \\
& & \big( \partial'_0\varphi(\xi')-A_0(\xi'_0,{\mathbf r}(\xi'_1)\big)
\nonumber\\
&+& \int d^2 \xi \int d^2 \xi'\,
\varphi(\xi) \, t_l(\xi_1) \partial_l M_{j}(\xi,\xi') t_j(\xi_1')\,
\big( \partial'_0\varphi(\xi')-A_0(\xi'_0,{\mathbf r}(\xi'_1)\big)
\nonumber\\
&+& \int d^2 \xi \int d^2 \xi'\,
t_l(\xi_1) A_l (\xi_0, {\mathbf r}(\xi_1)) M_{j}(\xi,\xi')
t_j(\xi_1')\,
\big(\partial'_0 \varphi(\xi')-A_0(\xi'_0,{\mathbf r}(\xi'_1)\big)
\nonumber\\
&-& \frac{i}{\kappa} \, \int d^2 \xi  \, \varphi(\xi) \, t_l(\xi_1) F_{0l}(\xi_0,{\mathbf r}(\xi_1))
\end{eqnarray}
where $F_{\alpha\beta}=\partial_\alpha A_\beta -\partial_\beta A_\alpha$, and
\begin{eqnarray}
M_{jk}(\xi,\xi')&=&  - \frac{i}{\kappa} \,\varepsilon_{jk}\, \langle
\xi_0,{\mathbf r}(\xi_1)| \frac{1}{-\partial^2}| \xi'_0,{\mathbf
r}(\xi'_1)\rangle \nonumber\\
M_j (\xi,\xi')&=&  \frac{i}{\kappa} \,\varepsilon_{jk}\, \langle \xi_0,{\mathbf r}(\xi_1)| \frac{\partial_k}{-\partial^2}| \xi'_0,{\mathbf r}(\xi'_1)\rangle \;.
\end{eqnarray}

It is straightforward to see that (\ref{eq:sc}) has the same transformation
properties as its fermionic equivalent. Indeed, all the terms in $S_{\mathcal
  C}$ except for the last one are invariant under  gauge transformations
restricted to the border:
\begin{equation}
A_\alpha(\xi_0, {\mathbf r}(\xi_1)) \,\to\,
A_\alpha(\xi_0, {\mathbf r}(\xi_1)) \,+\,
\partial_\alpha \omega (\xi_0, {\mathbf r}(\xi_1))
\end{equation}
if the scalar field is also transformed:
\begin{equation}
\varphi(\xi_0,\xi_1) \,\to\, \varphi(\xi_0,\xi_1) \,+\,
\omega (\xi_0, {\mathbf r}(\xi_1)) \;.
\end{equation}
It is clear that the last term in $S_{\mathcal C}$ does reproduce the chiral
anomaly, since under the previous gauge transformation:
\begin{equation}
\delta_\omega S_{\mathcal C} (\varphi,A) \;=\; -\frac{i}{\kappa} \,
\int d^2 \xi  \, \omega(\xi_0,{\mathbf r}(\xi_1))
\, t_l(\xi_1) F_{0l}(\xi_0,{\mathbf r}(\xi_1))\;,
\end{equation}
which is, of course, consistent with the result obtained from the fermionic
representation. The results agree when $\kappa = \frac{4 \pi}{e^2}$, which is
the proper value for the bosonized theory with `minimal' regularization.
\section{Linear wall}\label{sec:linear}
In this section we calculate the free fermion propagator and the
effective action for the particularly simple case of a linear
boundary, which we assume to be at $x_2 = 0$, with
\mbox{${\mathcal U} = \{(x_1,x_2): x_2 \geq 0 \}$}.

Let us first consider the free fermion propagator. Using $x_0$ and
$x_1$ as coordinates we write the free kernel ${\mathcal
K}_{linear}^{(0)}$ as
\begin{eqnarray}\label{eq:freek}
&&{\mathcal K}_{linear}^{(0)}(x_0,x_1; x_0',x_1') \;=\; \langle
x_0,x_1,0 |\frac{-\gamma_\mu \partial_\mu}{-\partial_\mu
\partial_\mu + m^2} {\mathcal P}_R|x_0',x_1',0\rangle \nonumber\\
&& =\; \langle x_0,x_1 | \frac{-\gamma_\nu \partial_\nu}{2
\sqrt{-\partial_\mu \partial_\mu + m^2}} {\mathcal P}_R
|x_0',x_1'\rangle \,,
\end{eqnarray}
where ${\mathcal P}_R = \frac{1 + \gamma_2}{2}$. Inserting this
expression into (\ref{freeprop}), after some algebra we obtain:
$$
\langle \psi(x) {\bar\psi}(y)\rangle \;=\; \langle
x|(\not\!\partial + m)^{-1}| y\rangle \;-\; 2 \, \int d^2x' \int
d^2x''
$$
\begin{equation}
\langle x| (\not\!\partial + m)^{-1}| x'_0,x'_1,0\rangle {\mathcal
P}_+ \langle x'_0,x'_1| V |x''_0,x''_1\rangle \, {\mathcal P}_-
 \langle x''_0,x''_1, 0 | (\not\!\partial + m)^{-1}| y \rangle \;.
\end{equation}
where ${\mathcal P}_{\pm} \equiv \frac{1 \pm \gamma_2}{2}$, and
\begin{equation}
V \;=\; \frac{\gamma_\mu \partial_\mu + m}{\sqrt{-\partial_\nu
\partial_\nu + m^2}}\, .
\end{equation}

Now we consider the evaluation of  the effective action
$\tilde{\Gamma}_f$, which is given by
\begin{equation}\label{eq:linear1}
\tilde{\Gamma}_f\;=\; - {\rm Tr} \ln {\mathcal K}_{linear}
\end{equation}
with
\begin{equation}\label{eq:linear2}
{\mathcal K}_{linear}(x_0,x_1; x_0',x_1') \;=\;
\langle x_0,x_1,0 |\frac{-\gamma_\mu D_\mu}{-(\not \!\! D)^2 + m^2}
{\mathcal P}_R|x_0',x_1',0\rangle \,.
\end{equation}
Note that, in the previous expression, the Dirac operator in the
denominator will in general depend on $A_2$, which does not commute with
$\partial_2$. Thus, in general, no simpler expression may be written for
(\ref{eq:linear2}) unless some simplifying assumptions are introduced.

Assuming that the $A_\alpha$'s are smooth functions of $x_2$ in the region
around $x_2=0$, the leading term in a $\partial_2$ derivative expansion is
$$
{\mathcal K}_{linear}(x_0,x_1; x_0',x_1') \;\simeq\;
$$
\begin{equation}\label{eq:linear3}
\langle x_0,x_1,0|
\frac{-\gamma_\mu D_\mu}{-(\gamma_\nu D_\nu)^2 - \partial_2^2 + e^2
A_2^2(x_0.x_1,0) +  m^2} {\mathcal P}_R |x_0',x_1', 0\rangle \,
\end{equation}
or,
$$
{\mathcal K}_{linear}(x_0,x_1; x_0',x_1') \;=\;
$$
\begin{equation}\label{eq:linear4}
\langle x_0,x_1 | \frac{-\gamma_\nu D_\nu}{2 \sqrt{-(\gamma_\mu
D_\mu)^2 + e^2 A_2^2(x_0.x_1,0)+ m^2}} {\mathcal P}_R
|x_0',x_1'\rangle \,,
\end{equation}
which is a sort of dimensional reduction of the original problem, although at
the expense of having to deal with a non-local theory.
This non local kernel may properly be called the effective Dirac operator for
the boundary modes, in a {\em microscopic\/} representation. It clearly shows
the well-known fact that the corresponding Dirac determinant contains gapless
excitations (as it should be~\cite{wen}) and also captures part of the non
locality which would have been lost if the $m \to \infty$ had been taken
beforehand.

The imaginary part of the effective action, on the other hand, can be borrowed
from the known result about the chiral fermion determinant in $1+1$ dimensions:
\begin{equation}\label{eq:imaglin1}
{\rm Im} {\tilde \Gamma}_f \;=\; \frac{e^2}{2\pi}
\int \, d^2x \, \frac{\partial \cdot  A}{\partial^2} \,
\epsilon_{\mu\nu} \partial_\mu A_\nu  \;,
\end{equation}
while for the real part we have:
\begin{equation}\label{eq:imaglin2}
{\rm Re} {\tilde \Gamma}_f (A) \;=\; - \frac{1}{2} {\rm Tr} \ln
\Big[\frac{-\gamma_\nu D_\nu}{2 \big(-(\gamma_\mu D_\mu)^2 + e^2
A_2^2(x_0,x_1,0)+ m^2\big)} \Big]\;.
\end{equation}

\section*{Acknowledgements}
C.D.F. thanks CONICET (PIP5531) for financial support. F.D.M.
acknowledges the warm hospitality of Centro At\'omico Bariloche,
where part of this work was done, and UBA, CONICET and ANPCyT for
financial support.


\begin{thebibliography}{bib}

\bibitem{review} M.\ Bordag, U.\ Mohideen and  V.\ M.\  Mostepanenko,
Phys.\ Rept.\ {\bf 353}, 1 (2001).

\bibitem{zeta}  S.\ Leseduarte and  A.\ Romeo, Ann.\ Phys.\  {\bf
250}, 448 (1996); E.\ Elizalde, F.\ C.\ Santos and  A.\ C.\ Tort,
Int.\ J.\ Mod.\ Phys.\  {\bf A18}, 1761 (2003); E.\ Elizalde, S.\
D.\ Odintsov, A.\ Romeo, A.\ A.\ Bytsenko and S.\ Zerbini {\em
Zeta function regularization techniques with applications}, (World
Scientific, Singapore, 1994).

\bibitem{beneventano} S.\ Leseduarte and  A.\ Romeo, Commun.\ Math.\
Phys.\ {\bf 193}, 317 (1998);
 C.\ G.\ Beneventano, M.\ De Francia, K.\ Kirsten and E.\ M.\ Santangelo, Phys.\ Rev.\
 {\bf D61}, 085019 (2000);
 M.\ Bordag and  K.\ Kirsten, Phys.\ Rev.\ {\bf D60}, 105019
 (1999).

\bibitem{kardar} H.\ Li and M.\ Kardar, Phys.\ Rev.\ {\bf A46}, 6490
(1992).

\bibitem{kardar2} T.\ Emig, A.\ Hanke, R.\ Golestanian, and M.\ Kardar,
Phys.\ Rev.\ Lett.\ {\bf 87}, 260402 (2001); ibidem Phys.\ Rev.\
{\bf A67}, 022114 (2003)

\bibitem{recent}T.\ Emig, R.\ L.\ Jaffe, M.\ Kardar and  A.\ Scardicchio,
Phys.\ Rev.\ Lett.\ {\bf 96}, 080403 (2006); M.\ Bordag,
hep-th/0602295.

\bibitem{wen}B.\ Blok and X.\ G.\ Wen,  Phys.\ Rev.\ {\bf B42}, 8133-8144
(1990).

\bibitem{zinn}See, for example, J.\ Zinn-Justin,
{\em Quantum Field Theory and Critical Phenomena}, Oxford Science
Publications, 4th. Ed., (2002).

\bibitem{Chodos:1974dm}
  A.~Chodos and C.~B.~Thorn,
  Phys.\ Lett.\ {\bf B53}, 359 (1974).

\bibitem{Milonni94} P.\ Milonni, {\em The Quantum Vacuum},
Academic Press, San Diego (1994).

\bibitem{b1}
  C.~P.~Burgess and F.~Quevedo,
  Nucl.\ Phys.\ B {\bf 421}, 373 (1994).

\bibitem{b2}
  A.~Kovner and P.~S.~Kurzepa,
  Phys.\ Lett.\ B {\bf 328}, 506 (1994).

\bibitem{b3}
  E.~H.~Fradkin and F.~A.~Schaposnik,
  Phys.\ Lett.\ B {\bf 338}, 253 (1994).

\bibitem{b4}
  D.~G.~Barci, C.~D.~Fosco and L.~E.~Oxman,
  Phys.\ Lett.\ B {\bf 375}, 267 (1996).
\end{thebibliography}
\end{document}